# TIME FLOW IN GRAPHENE AND ITS IMPLICATIONS ON THE CUTOFF FREQUENCY OF BALLISTIC GRAPHENE DEVICES

D. Dragoman[1] and M. Dragoman[2]

[1]Univ. Bucharest, Physics Dept., P.O. Box MG-11, 077125 Bucharest, Romania

[2]National Research and Development Institute in Microtechnology, Str. Erou Iancu Nicolae 126 A (32B), 077190 Bucharest, Romania

**Abstract**

This manuscript deals with time flow in ballistic graphene devices. It is commonly believed that in the ballistic regime the traversal time of carriers in gated graphene at normal incidence is just the ratio of the length of the device and the Fermi velocity. However, we show that the traversal time is much slower if the influence of metallic contacts on graphene is considered. Even the transmission at normal incidence becomes smaller than 1, contradicting yet another common belief. These unexpected effects are due to the transformation of Schrödinger electrons in the metallic contact into Dirac electrons in graphene and vice versa. As a direct consequence of these transformations, the ultimate performance of gated ballistic devices are much lower than expected, in agreement with experimental results.

---

a) Corresponding author email: danieladragoman@yahoo.com

**1. Introduction**

The time flow in graphene is an issue ignored in the huge literature accumulated on graphene. In ballistic graphene devices, such as transistors or multiple gated structures, the traversal time of carriers $\tau_0$ is calculated simply as the ratio between the length $L$ of the device (the gate length in field effect transistors) and the Fermi velocity $v_F$. Then, the cutoff frequency in the ballistic graphene device is defined as $f_c = (2\pi\tau_0)^{-1}$ [1]. However, if we apply this formula to experiments, the predicted cutoff frequency is of hundreds of GHz and even few THz in almost any graphene device. In the large majority of ballistic graphene devices such estimations are simply unrealistic. The aim of this paper is then to find an explanation of how time is flowing in graphene.

In an effort to refine the definition of traversal time in graphene, we introduce first a definition of $\tau_0$ inspired by the recently developed quantitative analogy between carrier propagation in graphene and that of holes in type II/III heterostructures [2]. The traversal time obtained in this way equals the expected value $\tau_0 = L/v_F$ only for normal incidence, but the results for oblique incidence do not differ significantly from this value except for the immediate neighborhood of the regions where electron propagation is forbidden; this issue is detailed in Section 2. In Section 3 we approached the problem in a much more radical way. More precisely, we considered the influence of contacts, in which electrons satisfy the Schrödinger equation. Then, the time flow in graphene is affected by the parameters of the contacts, such as the electron effective mass, and the traversal time reaches values in agreement with those obtained in recent experiments on high-frequency graphene transistors. The results obtained in Section 3 are based on the counterintuitive transformation of Schrödinger electrons in contacts to Dirac electrons in graphene and then back again, transformation that is certainly encountered in experiments but is difficult to describe mathematically, since it implies the conversion of a scalar wavefunction into two spinors, and vice versa. However, physical arguments can help solve this problem, the result being that time flow in graphene can be mathematically described including the presence of contacts. As a consequence, it is found that the cutoff frequency of ballistic graphene devices, in particular graphene transistors, is lower than that calculated with the simple formula $\tau_0 = L/v_F$, even for normal incidence. The results in this paper apply for ballistic charge carriers only. In real devices, back-injection from the drain and parasitic effects or scattering processes, which need a more complex treatment, only worsen the frequency performances of any devices, in particular graphene devices [3]. Therefore, a ballistic estimation provides, indeed, the ultimate cutoff frequency value.

**2. Time flow in gated graphene without electrical contacts**

The analogy between Dirac charge carriers in gated graphene regions and the two-band Kane carriers in type II/III semiconductor heterostructures [2] is based on the formal similarity of the evolution equation if the two spinor components in graphene, $\psi_1$ and $\psi_2$, are replaced by the envelope wavefunctions in the conduction and valence bands, $\psi_c$ and $\psi_v$, respectively, and if the Fermi velocity in graphene $v_F$ is replaced by the interband velocity matrix element $P$ between the conduction and valence bands in the semiconductor. This analogy allows, in a similar way as in type II/III heterostructures [4], the introduction of a velocity-group-based definition of the traversal time for electrons propagating along a distance $L$ in the $x$ direction in graphene as

$$\tau = \int_0^L \frac{dx}{v_g(x)} = \int_0^L \frac{\rho(x)dx}{J} \qquad (1)$$

where $v_g(x) = J/\rho(x)$ is the analog of group velocity in graphene, $\rho = |\psi_1|^2 + |\psi_2|^2$ is the probability density and $J = v_F(\psi_1 \psi_2^* + \psi_1^* \psi_2)$ is the probability current along $x$.

In order to compare $\tau$ as defined in (1) with $\tau_0$ and the frequency cutoffs in graphene transistors, we focus in this Section on a typical field-effect-transistor configuration consisting of a gated region of width $L$, through which electrons propagate along the $x$ axis. We allow for oblique incidence under an angle $\theta$ measured with respect to the $x$ axis and for a bias $V$ applied on the structure, approximating the linear potential drop across $L$ with a stepwise drop with value $eV/2$, as in [5]. The traversal time along the gated region determines the frequency cutoff of the transistor. Because $\psi_1$ and $\psi_2$ calculated as detailed in [5,6] depend on several parameters such as gate length $L$, the energy potential in the gate region $U_g$, the bias $V$ and the electron energy $E$, we study the influence of all these parameters on $\tau$. In Fig. 1 we have plotted the dependence on the incidence angle of the transmission and the traversal time normalized to $\tau_0$ for $E = 0.1$ eV, $V = 0.35$ V, $L = 50$ nm and $U_g = 0.2$ eV (solid line), 0.3 eV (dotted line) and 0.4 eV (dashed line). As in all other simulations in this paper, $T$ is represented with thin black curves and the normalized traversal time with thick gray curves, with the same line type (solid, dotted or dashed). All curves are symmetrical when $\theta$ is replaced by $-\theta$. From Fig. 1 it can be seen that, in all cases, for normal incidence ($\theta = 0$) the transmission equals 1, as expected, and $\tau = \tau_0$. A significant difference between $\tau$ and $\tau_0$ occurs only in the immediate neighborhood of the regions in which electron propagation is forbidden, and thus $T = 0$, when $\tau$ tends asymptotically to infinity; for $T = 0$, $\tau = \infty$ since electrons do not penetrate through the gate. These regions appear at oblique incidence whenever, depending of $U_g$, $E$ and $V$, the wavenumber of charge carriers in graphene becomes imaginary, situation that cannot be accommodated by the gapless band energy diagram in graphene (see [5] for a more detailed discussion of this situation and its implication on the electric transport in graphene). However, in almost all experiments electrons propagate normally (or very close to the normal) on the gated regions in field effect transistors, for example. Even if we take into account the contribution of electrons that depart from the normal with an angle up to, say, 5°, the average $\tau$ value is still very close to $\tau_0$ and the formula $\tau_0 = L/v_F$ seems to be supported by these simulations. The same conclusion can be drawn if other parameters are varied. In Fig. 2, for instance, the incidence-angle-dependence of $T$ and $\tau/\tau_0$ was represented for the same $E = 0.1$ eV, $L = 50$ nm and $U_g = 0.3$ eV, but different biases: $V = 0.25$ V (solid line), 0.35 V (dotted line) and 0.5 V (dashed line). A change in the energy of incidence charge carriers in graphene does not modify the conclusions, as can be seen from Fig. 3, where the transmission and normalized traversal time were plotted for $E = 0.1$ eV (solid line), 0.2 eV (dotted line) and 0.3 eV (dashed line) and the same $U_g = 0.3$ eV, $L = 50$ nm and $V = 0.35$ V. Although $L$ should not influence the position of the regions with $T = 0$, it could affect the traversal time, as shown in Fig. 4, where $T$ and $\tau/\tau_0$ as a function of the incidence angle were drawn with solid line for $L = 50$ nm, with dotted line for $L = 100$ nm and with dashed line for $L = 150$ nm, the other parameters being $E = 0.1$ eV, $U_g = 0.3$ eV and $V = 0.35$ V. As a general conclusion from all these simulations, in the neighborhood of forbidden regions for charge carrier propagation in graphene $\tau/\tau_0$ seems to have a larger slope as the decrease of $T$ is sharper, but near normal incidence $\tau \cong \tau_0$ irrespective of the parameters used in computations.

### 3. Time flow in gated graphene with electrical contacts

The results of the previous section are not encouraging if the cutoff frequencies of ballistic graphene transistors are to be compared with simulations. In such transistors the group velocity of carriers, $v_g$, is almost equal to the drift velocity $v_d$ calculated using only transistor geometrical data and its dc parameters as $v_g \cong v_d = L g_m / C_g$, where $L$ is the gate length, $C_g$ is the gate capacitance, and $g_m$ is the graphene transconductance. In general, $v_d < v_F$ (it is almost equal to $v_F/2$ in [1]), and the corresponding traversal time should satisfy the relation $\tau > \tau_0$. This relation must hold at (or very near) normal incidence, which is the case in practice. To explain these facts in a ballistic theory of graphene, one should include the electrical contacts in the analysis. After all, the electrons in the source and drain of the transistor (metallic electrodes) satisfy the Schrödinger equation, transforming into Dirac-like electrons and back again while traversing the graphene sheet (channel) between the electrodes. We consider in this paper only ohmic contacts, because a recent experimental work [7] demonstrates that a suitable sample treatment can improve the contact resistance of almost all metals, rendering Schottky contacts into almost ohmic-like contacts. As a consequence, the selection of metal is of little importance. From a cutoff frequency point of view, ohmic contacts are ideal.

Therefore, a mathematical method to describe the transformation of Schrödinger electrons into Dirac electrons and vice-versa is required. This transformation takes place at the electrode/graphene interface; we consider in this section only normal incidence. The scalar wavefunction of electrons in the source and drain (regions 1 and 3, respectively) is taken to be

$$\Psi = \begin{cases} A_1 \exp(ik_1 x) + B_1 \exp(-ik_1 x), & x \leq 0 \\ A_3 \exp(ik_3 x), & x \geq L \end{cases} \quad (2)$$

with $k_1 = (2mE)^{1/2}/\hbar$, $k_3 = [2m(E+eV)]^{1/2}/\hbar$, and $m$ the effective electron mass in the metallic electrodes, while in the gated graphene channel, for $0 < x < L$, we have

$$\Psi = \begin{pmatrix} \psi_1 \\ \psi_2 \end{pmatrix} = \begin{pmatrix} A_2 \exp(ik_2 x) + B_2 \exp(-ik_2 x) \\ s_2 [A_2 \exp(ik_2 x) - B_2 \exp(-ik_2 x)] \end{pmatrix} \quad (3)$$

with $k_2 = (E - U_g + eV/2)/\hbar v_F$. At the electrode side of the interfaces the wavefunctions and its $x$ derivative normalized to $m$ must be continuous, while at the graphene side of the interfaces the two spinor components must be continuous. Then, we impose the following boundary conditions at the $x = 0$ and $x = L$ interfaces, respectively:

$$A_1 + B_1 = A_2 + B_2, \quad \frac{k_1}{m}(A_1 - B_1) = \frac{v_F}{\hbar}s_2(A_2 - B_2), \tag{4a}$$

$$A_2 e^{ik_2 L} + B_2 e^{-ik_2 L} = A_3 e^{ik_3 L}, \quad \frac{v_F}{\hbar}s_2(A_2 e^{ik_2 L} - B_2 e^{ik_2 L}) = \frac{k_3}{m}A_3 e^{ik_3 L}. \tag{4b}$$

The form of the boundary conditions can be justified by considering that, if $A_j$, $B_j$, $j = 1,2,3$ are the amplitudes of the forward and backward plane wave components, the wavefunction in regions 1 and 3 is similar to $\psi_1$, while the derivatives of the wavefunction in the electrodes have the same form as $\psi_2$. Moreover, the constant $v_F/\hbar$ is required from dimensionality considerations. The boundary conditions in (4) also guarantee that the current probability is conserved across the structure, result that can be checked by a straightforward calculation; in electrodes, $J_j = (\hbar/2mi)[\Psi_j^*(d\Psi_j/dx) - \Psi_j(d\Psi_j^*/dx)]$, $j = 1,3$.

The influence of electrodes on electron transmission and the traversal time in graphene is dramatic. In this case the transmission probability is defined as $T = (k_3/k_1)|A_3/A_1|^2$. A plot of the energy dependence of $T$ (thin black lines) and the normalized traversal time (thick gray lines, of the same type as the corresponding $T$) for $V = 0.35$ V, $L = 50$ nm, $m = m_0$ with $m_0$ the free electron mass, and $U_g = 0.2$ eV (solid line), 0.3 eV (dotted line) and 0.4 eV (dashed line), as displayed in Fig. 5, shows that when the electrical constants are taken into account, the transmission probability at normal incidence is not equal to 1 as a rule. Moreover, the three corresponding traversal times are superimposed in Fig. 5, which means that $\tau$ does not depend on $U_g$, and $\tau/\tau_0$ is significantly greater than unity. This means that electrons propagate slower, i.e. with an effective drift velocity $L/\tau$ smaller than $v_F$, even if the transport is still considered as ballistic. The $T<1$ values in Fig. 5 for normal incidence are caused by the mismatch between electrodes and graphene due to the different evolution laws that electrons obey in the two situations. The fact that the transmission probability has an oscillatory behavior with $E$ is an indication of interference taking place in graphene between the forward and backward propagating plane wave components of the wavefunction. From Fig. 5 it also follows that the traversal time decreases as $E$ increases, the more energetic charge carriers traversing the structure faster. Although $\tau$ does not depend on $U_g$, it does depend on $V$, as can be seen from Fig. 6, and is found to be independent of $L$, as suggested by Fig. 7. The parameters used for the simulations in Fig. 6 were: $U_g = 0.3$ eV, $L = 50$ nm, and $E = 0.1$ eV (solid line), 0.2 eV (dotted line) and 0.3 eV (dashed line), whereas in Fig. 7 we considered $U_g = 0.3$ eV, $V = 0.3$ V, and $E = 0.1$ eV (solid line), 0.15 eV (dotted line) and 0.2 eV (dashed line). Figure 6 suggest that $\tau$ decreases as the bias increases, at least for the parameters used in the simulation. Figure 7, which shows that the effective drift velocity $L/\tau$ is independent of $L$, is in agreement with the experimental results in [1]. Moreover, the drift velocity in [1], which is approximately $v_F/2$, value that corresponds to $\tau/\tau_0 \cong 2$, is within the range of simulated traversal times in Figs. 6 and 7; as $\tau/\tau_0 = \tau v_F/L$, Fig. 7 also shows that the traversal time is proportional to the channel length $L$ of graphene-based transistors while the cutoff frequency is inversely proportional to $L$, as verified by measurements [1]. So, the experimental data on cutoff frequencies of real graphene-based transistors can be explained by including the metallic contacts in the analysis, even if normal incidence and the ballistic transport regime are assumed. Our simulations suggest that higher cut-off frequencies can be obtained in shorter devices traversed by energetic charge carriers (far from Dirac point) under large bias conditions.

## 4. Conclusions

Using the analogy between Dirac-like charge carriers in graphene and Schrödinge-like carriers in type II/III semiconductor heterostructures, we have defined the traversal time in graphene. We have demonstrated that this traversal time cannot be described by the simple formula $\tau_0 = L/v_F$, in particular in gated graphene regions. The traversal time is equal to $\tau_0$ only for normal incidence in contact-less graphene structures, but increases significantly near the regions where electron propagation is forbidden. However, close to normal incidence the traversal time remains very close to $\tau_0$, so that the experimental frequency cutoff values of graphene-based transistors cannot be justified by a ballistic theory of charge carriers unless the electrical contacts are taken into account. The passage of electrons from the source contact to graphene and from graphene to the drain contact implies mathematically a transformation from the Schrödinger equation to the Dirac equation and back again, which seems impossible to perform due to the different forms of these equations. However, we have proposed such a transformation, which conserves the current probability throughout the structure and leads to interesting results. One of these is that the transmission is no longer unity at normal incidence due to the mismatch between graphene and contacts. Another consequence is that the traversal time is significantly greater than $\tau_0$ and the effective drift velocity is significantly smaller than $v_F$ even in the ballistic transport regime. The last conclusion is supported by experimental results. So, despite the initial enthusiasm to reach very high THz frequencies with graphene transistors, our results show that the expectations should be much moderate, even in ballistic devices.

**Figure captions**

Fig. 1 The dependence on the incidence angle of $T$ (thin black curves) and $\tau$ (thick gray curves) for $E = 0.1$ eV, $V = 0.35$ V, $L = 50$ nm and $U_g = 0.2$ eV (solid line), 0.3 eV (dotted line) and 0.4 eV (dashed line).

Fig. 2 The dependence on the incidence angle of $T$ (thin black curves) and $\tau$ (thick gray curves) for $E = 0.1$ eV, $L = 50$ nm, $U_g = 0.3$ eV, and $V = 0.25$ V (solid line), 0.35 V (dotted line) and 0.5 V (dashed line).

Fig. 3 The dependence on the incidence angle of $T$ (thin black curves) and $\tau$ (thick gray curves) for $U_g = 0.3$ eV, $L = 50$ nm, $V = 0.35$ V and $E = 0.1$ eV (solid line), 0.2 eV (dotted line) and 0.3 eV (dashed line).

Fig. 4 The dependence on the incidence angle of $T$ (thin black curves) and $\tau$ (thick gray curves) for $E = 0.1$ eV, $U_g = 0.3$ eV, $V = 0.35$ V and $L = 50$ nm (solid line), 100 nm (dotted line) and 150 nm (dashed line).

Fig. 5 Energy dependence of $T$ (thin black curves) and $\tau$ (thick gray curves) for normally incident electrons emitted from contacts with $m = m_0$ for $V = 0.35$ V, $L = 50$ nm, and $U_g = 0.2$ eV (solid line), 0.3 eV (dotted line) and 0.4 eV (dashed line).

Fig. 6 Bias dependence of $T$ (thin black curves) and $\tau$ (thick gray curves) for normally incident electrons emitted from contacts with $m = m_0$ for $U_g = 0.3$ eV, $L = 50$ nm, and $E = 0.1$ eV (solid line), 0.2 eV (dotted line) and 0.3 eV (dashed line).

Fig. 7 Dependence of $T$ (thin black curves) and $\tau$ (thick gray curves) on $L$ for normally incident electrons emitted from contacts with $m = m_0$ for $U_g = 0.3$ eV, $V = 0.3$ V, and $E = 0.1$ eV (solid line), 0.15 eV (dotted line) and 0.2 eV (dashed line).

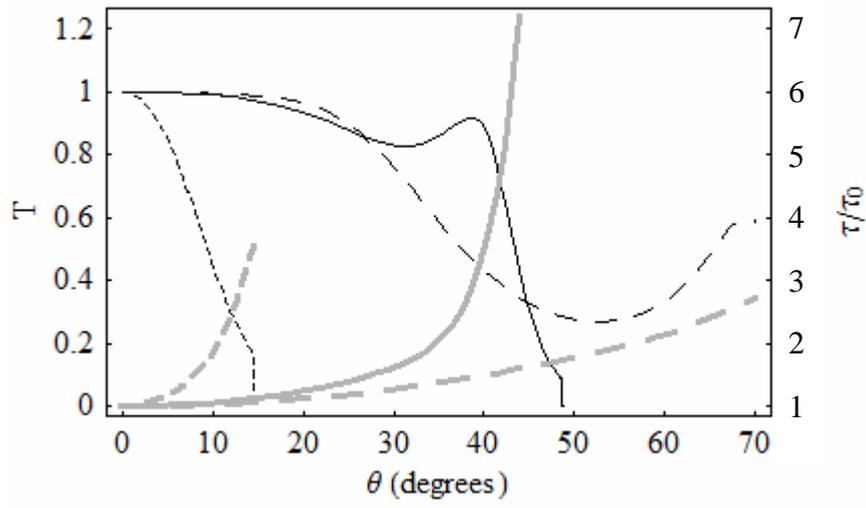

Fig. 1

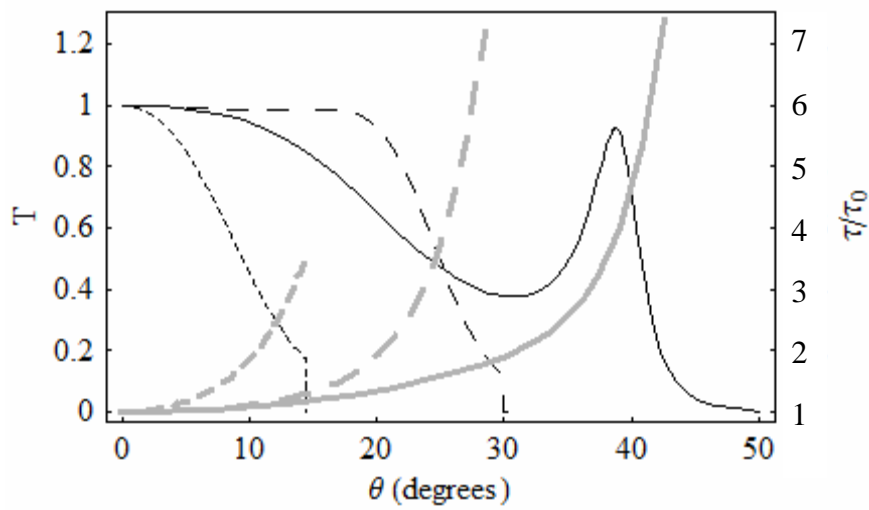

Fig. 2

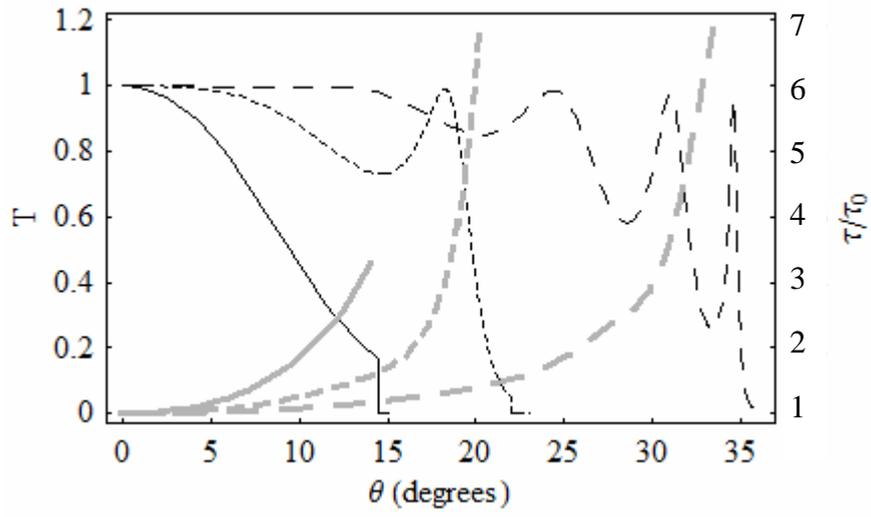

Fig. 3

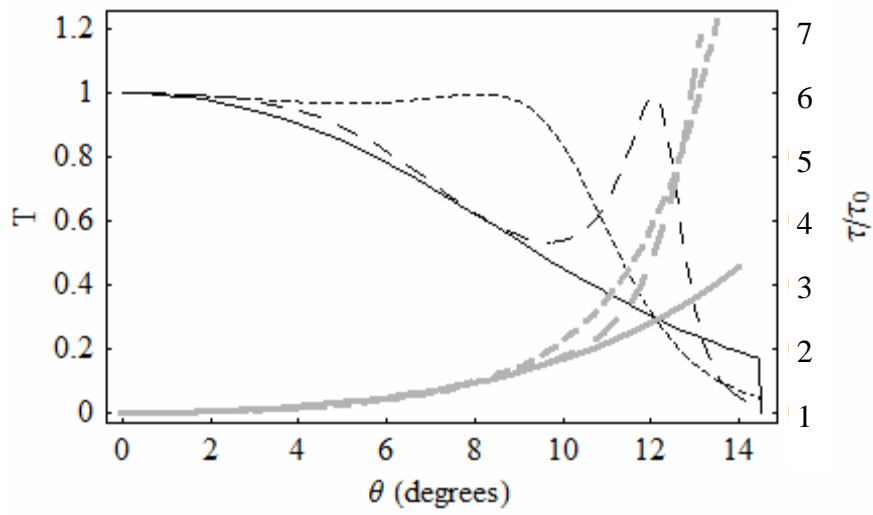

Fig. 4

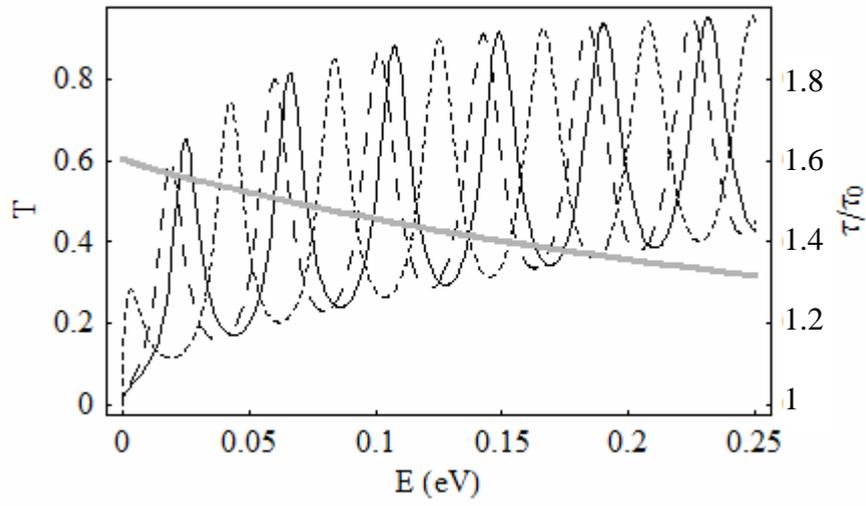

Fig. 5

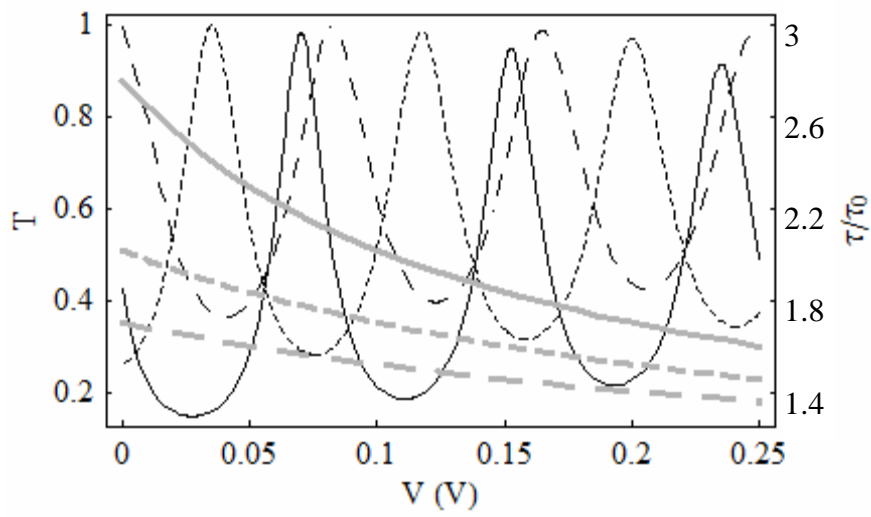

Fig. 6

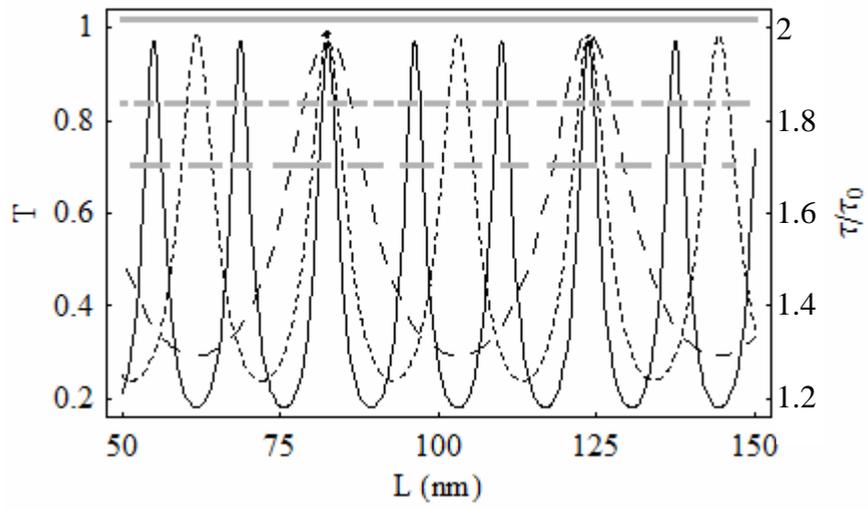

Fig. 7